\documentstyle[12pt,aaspp4]{article}

\begin{document}

\title{Using Quadruple Lenses to probe the Structure of the Lensing Galaxy}

\author{Hans J. Witt}
\affil{Astrophysikalisches Institut Potsdam, An der Sternwarte 16, 
       14482 Potsdam, Germany}
\authoremail{hwitt@aip.de}
\begin{abstract}
We show here that quadruple lenses can be useful laboratories to
probe whether the potential of the lensing galaxy is purely
elliptical or whether an additional distortion is present in the
deflector plane. 
For this test we only have to know the relative image positions 
of the quadruple lens system and
the (relative) center of light position of the lensing galaxy.

Furthermore we introduce new methods which easily allow us
to determine
the location (rotation angle relative to the image positions) 
of the major axis of the lensing galaxy.
In due course we can determine
the parity of the four images as well.
We apply these methods  to the 8 currently
known quadruple lenses
and find that in the case of MG 0414+0534, CLASS 1608+656
and HST 12531--2914 it is impossible to accommodate
the relative image positions and the galaxy position with any elliptical
potential whereas the other five cases can be described very well
with a simple elliptical potential.
This method will have important impacts for $\chi^2$-fits and
the reconstruction of galaxy models for quadruple lenses.
\end{abstract}

\section{INTRODUCTION}

Quadruple lenses are very special configurations where a
quasar is located behind the center of a lensing galaxy.
Only such an arrangement can plausibly reproduce
such image configurations.
Currently there are already eight quadruple lenses
known in the optical band (see Keeton \& Kochanek 1996 and references
therein, hereafter referred to as KK96).

For most of the quadruple lenses the image configuration
can be reconstructed by using a simple galaxy model,
e.g. an elliptical potential or a circular potential plus shear
(see e.g. Kochanek 1991, Wambsganss \& Paczy\'nski 1994, Witt, Mao \& Schechter
1995).
However, in the case of MG0414+0534 it seemed rather difficult to find
an adequate model which has a small number of free parameters
and a $\chi^2$ which is close to one per degree of freedom
(cf. Falco et al. 1996).
I show in this {\it letter} that for 
the reconstruction of some quadruple lenses
an elliptical potential or a circular potential plus external shear
can not provide a satisfactory solution independent on the number of free 
parameters for the $\chi^2$-fit.

\section{PROPERTIES OF ELLIPTICAL POTENTIAL}

A very commonly used potential for galaxy models  in the gravitational lensing
literature is either an elliptical potential like
the (elliptical) power-law potential
(see e.g. Kassiola \& Kovner 1993 and references therein) or
just a circular potential plus an external shear
(see e.g. Kochanek 1991, Wambsganss \& Paczy\'nski 1994).
I show here that both potential have remarkable properties 
concerning the possible position of the lensing galaxy. 

We assume a given two-dimensional elliptical potential in the deflector plane
in the form $\psi( x^2 +y^2/q^2)$ where $q$ $(0<q\leq 1)$ is the axis ratio
of the elliptical potential. 
The (projected) surface mass distribution is then given by
$\Delta \psi = 2 \kappa(x,y)$, where 
$\kappa(x,y)=\Sigma(x,y)/\Sigma_{\rm crit}$
is expressed in units of the critical density $\Sigma_{\rm crit}$
which depends on the distances to the deflector and the source
(cf. Schneider, Ehlers \& Falco 1992).
The two-dimensional deflection angle is then simply given by
the derivatives of the potential,
{\boldmath $\alpha$}$=${\boldmath $\nabla$}$\psi$. 

Considering now the two dimensional lens equation in
normalized coordinates, we can write
\begin{eqnarray} 
\xi &=& x + \gamma x
         - \frac{\partial \psi((x-x_G)^2+(y-y_G)^2/q^2)}{\partial x} 
         = x +\gamma x - \frac{\partial \psi(r_e)}{\partial r_e} 2 (x-x_G) 
 \label{xi} \\  \label{eta}
\eta &=& y -\gamma y
       - \frac{\partial \psi((x-x_G)^2+(y-y_G)^2/q^2)}{\partial y }
      = y - \gamma y
    -\frac{\partial \psi(r_e)}{\partial r_e} \frac{2(y-y_G)}{q^2} 
\end{eqnarray}
where $(x_G,y_G)$ is the position of the center of the lensing galaxy,
$\gamma$ is the shear which is supposed to act along the $x$-axis 
(major axis, $\gamma > 0$) or along the $y$-axis (minor axis, $\gamma < 0$)
and $r_e=(x-x_G)^2+(y-y_G)^2/q^2$. (Note that a purely elliptical potential
$(\gamma=0)$ and a circular potential plus shear $(q=1)$
are contained as special cases in the equations).

Through this paper we are choosing a special coordinate system
where the source is at the origin $(\xi=\eta=0)$. Now we can move each of the
the last term of eqs.(\ref{xi}) and (\ref{eta}) on the left side of the
equations and divide eq.(\ref{eta}) by eq.(\ref{xi}).
Assuming the potential 
yields at least 4 images (as observed for the case of quadruple lenses)
we find the following relation for the image positions in this 
particular coordinate system:
\begin{equation}
\label{yixi} \frac{y_i(1-\gamma)}{x_i(1+\gamma)} 
= q^{-2} \frac{(y_i-y_G)}{(x_i-x_G)}
\quad \mbox{for} \quad i=1,...,4,....
\end{equation}
Combining these four equations,
we can now first eliminate $q$ and $\gamma$ and  then further
$x_G$ by assuming that we know
the image positions (for this particular coordinate system).
Eliminating first $q$ and $\gamma$ we obtain
\begin{equation}
\label{identG}
x_G y_G [ x_1 y_2 - x_2 y_1] + x_2 y_2 [x_G y_1 -x_1 y_G]+ x_1 y_1
[x_2 y_G - x_G y_2] =0 \quad . 
\end{equation}
where the index 1 or 2 may be replaced by an index of $i=3,4,5$, respectively.

Further elimination of $x_G$ yields
that $y_G$ factorizes automatically out of the equation by this process. 
In other
words $x_G$ can not be determined independently with the
third image position.
The final relation for 3 image positions is given by
\begin{equation}
\label{identxy}
x_3 y_3 [ x_1 y_2 - x_2 y_1] + x_2 y_2 [x_3 y_1 -x_1 y_3]+ x_1 y_1
[x_2 y_3 - x_3 y_2] =0 
\end{equation}
which has the same structure as eq.(\ref{identG}). 
In this relation each index may be replaced by the index of another image, 
i.e. by the index
of the fourth or fifth image.
The problem is that usually only the relative image positions are observed
(the first image is placed at the origin)
and the position of the source is unknown. 
However, we can relate the relative image positions 
with this particular coordinate system. 
The two coordinate systems are related by a shift in the 
$x$ and $y$-direction and by a rotation of an angle $\theta$.
We note that eq.(\ref{identG}) and eq.(\ref{identxy}) are a function
of the form $y_i= a_1 x_i/(1+a_2 x_i)$ for $i=1,...,4, G$, whereby
$a_1$ and $a_2$ are coefficients which are determined by two
other image positions. 
If we transform now this kind of equation into a coordinate
system with relative image positions (shifted and rotated coordinates
which are denoted with a prime)
we obtain the following equation $(x_1'=y_1'=0)$:
\begin{equation}
\label{general} c_1 {x'}^2 + c_2 x'y' -c_1 {y'}^2 + c_3 x' + y' =0 
\end{equation}
This equation is the main result of the paper.
It is the general equation which yields the possible 
position of the lensing galaxy and the
image positions assuming an elliptical potential plus shear as given in
the lens equations (\ref{xi}) and (\ref{eta}).
The coefficients $c_1$, $c_2$ and $c_3$ are uniquely determined by the three
relative image positions of the quadruple lens $(x_i', y_i')$ for $i=2,3,4$ 
(or by two relative image positions
and one relative galaxy position) and can be obtained by using for example
Cramer's rule. 
Equation (\ref{general}) is almost comparable with an equation of a hyperbola,
but here we have an additional mixed term $x'y'$.
We emphasize here that a possible central fifth image 
(or any additional image) must
be located on this curve as well. Also the (relative) source position 
is located
on this curve assuming the lens equation as given 
in eq.(\ref{xi}) and (\ref{eta}). We shall keep this in mind for
the discussion below.
In Figure 1 two examples of quadruple lenses are shown
where the line 
of eq.(\ref{general}) is indicated. 

\section{The Major Axis of the Lensing Galaxy}

As a corollary we can now determine the rotation angle of the major axis
of the elliptical potential (or the direction of the shear for $q=1$, 
see \S 2)
relative to the right ascension $(\Delta\alpha)$ and declination offset
$(\Delta \delta)$  of the (relative)
image positions. If we rotate eq.(\ref{identG}) by an
angle $\theta$, this would yield the general form of eq.(\ref{general}).
Taking this into account we obtain
\begin{equation}
\tan (2\theta) =  -2 \frac{c_1}{c_2}   \label{angle}
 = \frac{2 \sum\limits_{i=2}^4  
x_i' y_i' [ x_{[i+1]}' y_{[i+2]}' - x_{[i+2]}' y_{[i+1]}' ]  }{ 
\sum\limits_{i=2}^4 ( {x_i'}^2 -{y_i'}^2 ) 
[ x_{[i+1]}' y_{[i+2]}' - x_{[i+2]}' y_{[i+1]}']  }
\end{equation}
where the square brackets at the index are defined 
as $[5]=2$, $[6]=3$ and $[i]=i$ else.
This equation contains the 3 relative image positions (which can be observed
in the case of quadruple lenses). 
Therefore we can determine the rotation angle $\theta$ 
of the major axis of the lensing galaxy relative to the observed image
positions of a quadruple lens. However, this relation does not give
us a unique angle. It only gives us the rotation angle modulo $90^\circ$. 
Therefore we have still two choices, how the major axis can be located
(i.e. rotated).
To obtain an unique solution we need further information about the parity
of the images. 
We now show a way how the parity of the images can be determined 
and hence how to uniquely determine the rotation angle, if the
flux ratios of the images are known.

\subsection{Parity of the Images} 

For the case that the 
center of the galaxy is very close to the line
of sight ($x_G \ll 1'', y_G\ll 1''$) we expect that the images of
negative parity are located close (``parallel'') to the major axis of
the lensing galaxy and that
the images
of positive parity  are located further away (``perpendicular'') to the major
axis of the lensing galaxy (cf. Fig. 1).
This must be the case because the critical curves
have elliptical form with the same alignment of the major axis.
Since the images of a quadruple lens are almost located on a circle
(Einstein ring) the two images along the major axis must be located
inside the critical curve at the region of negative parity.
Without any proof we state here that inside the region where the maximum
number of $n$ images may appear
\begin{equation}
\label{mupi} \sum_{i=1}^n p_i \mu_i \geq 1 
\end{equation}
does hold. $p_i$ denotes the parity of the images $(p_i=+1$ or $-1)$ 
and $\mu_i$ is the magnification of the images. This relation is well
motivated because it can be shown that for every (elliptical) power-law galaxy
as studied by Evans (1994) this relation is true (Witt 1996, in preparation).
Especially if the mass of the deflector is more centrally concentrated
the left side of the relation is close to one and for flatter mass
distribution the relation on the left side becomes larger.
Dividing eq.(\ref{mupi}) by the magnification of one of the images yields
that the sum of the flux ratios times the parity must be always positive.
If the flux ratios are observed we can uniquely determine the rotation angle
$\theta$ by testing relation (\ref{mupi}) providing that the flux ratios
are not heavily influenced by microlensing. This is certainly not the case
in the radio band where the source size is too large to obtain
significant variations caused by microlensing (cf. the discussion in
Witt et al. 1995).

In the case that two bright images are close together like
in PG\,1115+080 and MG\,0414+0534 (see KK96)
we can use the relation (\ref{mupi}) to determine the parity immediately.
Since the two bright images have opposite parity and almost equal
brightness they must cancel each other in the sum of eq.(\ref{mupi}).
The third brightest image must have positive parity and consequently the
fourth more fainter image must have negative parity to assure
that the sum is positive in eq.(\ref{mupi}). The four images of the lens
form always almost a circle whereby the parity should alternate
along the circle. This determines the parity of the two bright images.
We note that a possible central fifth image must have positive parity.

\section{DISCUSSION}

We showed here that for any elliptical potential or any circular
potential plus external shear that produce four images
the image positions and the
position of the lensing galaxy are related.
They must be located on certain lines
which are determined by eq.(\ref{general}) (cf. also Figure 1). 
In addition it is shown
that the rotation angle of the major axis can be determined
from the relative image positions of a quadruple lens. In Table 1
we present the rotation angle $\theta$ and the minimum
discrepancy $\Delta_{\min}$ (in arcsec)
of the observed galaxy position to the curve of eq.(\ref{general})
for the 8 currently known quadruple lenses.
In three cases, i.e. for MG 0414+0534, CLASS 1608+656 and HST 12531--2914,
we find that the discrepancy $\Delta_{\min}$ is much larger $(>3\sigma)$
than the claimed measurement error of the galaxy position.
For these cases it is impossible to achieve a good $\chi^2$-fit
by using an elliptical potential and to try to accommodate 
the galaxy and the image positions simultaneously.

We would like to emphasize here that the lines as shown in Fig. 1 are a 
unique property of the elliptical potential. 
As far as we know they do not exist for other potential.
Only in the case of elliptical potential (circular potential plus shear)
we find lines in the parameter space which gives the possible
location of the galaxy while in other cases we find isolated discrete points
in the parameter space which yields us the position of the galaxy.
As an example
we can start with an potential in the form $\psi(x^4+y^4/q^4)$.
Such an potential would be boxier than an elliptical potential.
Doing the steps as described in \S 2 we end up with an
equation in the form $f(\{x_i',y_i',i=2,3,4,G\}, \theta)=0$.
In this case the galaxy position does not simply factorize out of the equation.
The consequence is that we need for this case three relative image positions 
plus
the relative galaxy position to determine the rotation angle $\theta$.
But for this case we are able to determine the relative position of the source
and $q$, as well. Surprisingly we found one or more (discrete) solutions
for each quadruple lens in Table 1 (except for H 1413+117 where the galaxy
position has not been observed). This means we can not exclude automatically
a potential in the form $\psi(x^4+y^4/q^4)$ for any known quadruple lens.
The situation is very different for elliptical potential.
The relative source position is located somewhere on the line as indicated
in Fig. 1. Thus we are not able to determine $q$ from
the image positions or galaxy position when we use an elliptical
potential. 
 
The consequences are very severe when we choose an elliptical potential
to use it for a $\chi^2$-fit of a quadruple lens. 
If we use only the relative image and galaxy position
for the $\chi^2$-fit of an elliptical potential (or circular potential
plus shear) we find that the $\chi^2$ (nearly) degenerates in $q$ ($\gamma$).
Consequently we find models with higher magnification
for $q$ closer to one (or for smaller $\gamma$), but the
$\chi^2$-fit does not show significant differences. This phenomenon 
is confirmed by our numerical calculation using an 
elliptical power-law potential by Evans (1994),
(cf. also Wambsganss \& Paczy\'nski 1994 for a circular potential plus shear).

The recommendations are as follows:
If only the relative image and galaxy positions of a quadruple
lens are known an elliptical potential or a circular potential plus shear 
should not be used for the $\chi^2$-fit because it tends to degenerate.
(This is also true for elliptical mass distribution if $0 \ll q <1$.
In this case elliptical mass distribution behave similar 
like elliptical potential). In this case it is better to use
for example more boxier potentials or potentials which deviate considerably
from elliptical potential.
If the flux ratios are known in addition, an elliptical potential
may be used for the $\chi^2$-fit when (a) the galaxy position
is close to the line of eq.(\ref{general}) and when (b) the flux ratios
are incorporated appropriately in the $\chi^2$-fit.
If we add an external shear $(\gamma_1,\gamma_2)$ to an elliptical potential
(which acts not along the major axis) we obtain a whole set of possible 
solutions $(\gamma_1,\gamma_2)$ 
for the external shear which is able to accommodate the
3 relative image position and the relative galaxy position, as well. In this
case we need the flux ratios in addition to determine  
the two components of the external shear.

More exact determination of the relative image positions,
the relative galaxy positions and the flux ratios is needed to discriminate
an elliptical potential or another potential for the observed
quadruple lenses. This is especially the case for 2237+0305
where the discrepancy of the galaxy position is very 
close to the 3$\sigma$-level (cf. Table 1).  
In addition this method can help to narrow considerably the effort
to search for the possible location 
of the lensing galaxy (especially for the case of H 1413+117) and a possible
central fifth image.
Further this test  might be useful to resolve the confusing
situation of 2016+112 (see e.g. Garrett et al. 1994 and KK96)
 and it can be tested whether 2016+112
consist out of two, three or eventually more images and it can be tested
where the lensing galaxy might be located.

\acknowledgments

I am grateful to Emilio Falco and Joseph Leh\'ar for encouragement 
and discussion and I like to thank Joachim Wambsganss for comments on the
manuscript.
This work was supported by a postdoctoral grant of the Deutsche
Forschungsgemeinschaft (DFG) under Gz. Mu 1020/3-1.

\clearpage

%
%

\begin{table}[ht]
\centerline{TABLE 1}
  {
  \begin{center}
\parbox {14.0cm}{Currently Known Quadruple Lenses}
\begin{tabular}{lrcll}
    \multicolumn{5}{c}{}\\
    \hline \hline
    &&&&\\
  Object & $\theta [^{\circ}]$ & $\Delta_{\min} ['']$ & 
$\sigma_{\rm galaxy}['']$ & Reference \\
    &&&&\\
    \hline
    &&&&\\
2237+0305 & --23.2 & 0.014 & 0.005 & Crane et al.\ 1991     \\
PG 1115+080 & --24.2 & 0.062 & 0.05 & Kristian et al.\ 1993  \\
MG 0414+0534 & --8.9 & 0.144 & 0.03 & Ellithorpe 1995, Schechter \& Moore 1993
 \\
CLASS 1608+656 & --21.5 & 0.129 & 0.01 & Myers et al.\ 1995, Schechter 1995  \\
B 1422+231 & 36.1 & 0.032 & 0.05 & Patnaik et al.\ 1992, Yee \& Ellingson 1994
  \\
H 1413+117 & 21.8$^*$ & --- & --- & Schechter 1995, Kayser et al. 1990  \\
HST 14176+5226 & --41.0 & 0.068 & 0.03 & Ratnatunga et al.\ 1995, 1996   \\
HST 12531--2914 & 18.6$^*$ & 0.224 & 0.03 &  Ratnatunga et al.\ 1995, 1996 \\
&&& \\
     \hline
&&&\\
  \end{tabular}

\parbox {14.0cm}{Note.--- $\theta$ is the rotation angle
of the major axis of the lensing galaxy relative to the observed image position
(cf. Figure 1) and $\Delta_{\min}$ 
is the minimum distance (in arcsec) of the observed galaxy position to the
curve of eq.(\protect\ref{general}), (cf. solid line in Figure1). 
$\sigma_{\rm galaxy}$ denotes the claimed
measurement error of the relative position of the lensing galaxy
(see reference). The asterisk denotes the cases where $\theta$ could
not be determined unambiguously because of  different observed flux ratios in 
different colors (cf. \S 3).
For H1413+117 no galaxy position was identified yet.}

  \end{center}
}
\end{table}

\clearpage
 
%
%

\begin{figure}
\caption{Relative image and galaxy positions are shown
for two observed quadruple lenses
(for references see Table 1). The solid
line indicates the region where the source position, the position
of the galaxy and the position of the central image may be located
as obtained by eq.(\protect\ref{general}) (see text).
The dashed line indicates the location of the major axis of the
lensing galaxy as obtained by eq.(\protect\ref{angle}).
Note that the images close to the dashed line have negative parity
(see \protect\S 3.1).
}
\end{figure}


\end{document}